\documentclass[]{proarticle}
\usepackage{multicol}
\usepackage{graphicx}
\usepackage{amssymb,mathrsfs,amsfonts}
\usepackage{hyperref}

\newcommand{\beq}{\begin{equation}}
\newcommand{\eeq}{\end{equation}}
\newcommand{\vp}{\vphantom}
\def\bea{\begin{eqnarray}}
\def\eea{\end{eqnarray}}
\begin{document}
\title{A note about fermionic equations of $AdS_4\times\mathbb{CP}^3$ superstring}
\author{D.V. Uvarov}
\maketitle
\address{NSC Kharkov Institute of Physics and Technology, Kharkov, 61108, Ukraine.}
\eads{d$\_$uvarov@hotmail.com}
\begin{abstract}
It is proved that when 8 fermions associated with the supersymmetries broken by the $AdS_4\times\mathbb{CP}^3$ superbackground are gauged away by using the $\kappa-$symmetry corresponding equations obtained by variation of the $AdS_4\times\mathbb{CP}^3$ superstring action are contained in the set of fermionic equations of the $OSp(4|6)/(SO(1,3)\times U(3))$ sigma-model.
\end{abstract}

%% \MSC 78A35 \sep 78-05 \sep 81V45

\begin{multicols}{2}
\section{Introduction}

The motivation for studying IIA superstring theory on the
$AdS_4\times\mathbb{CP}^3$ superbackground comes from the
$AdS_4/CFT_3$ correspondence in the formulation by Aharony,
Bergman, Jafferis and Maldacena \cite{ABJM} conjecturing that in
the limit $k^5\gg N\gg 1$ it describes gravity dual of  the $D=3$
$\mathcal N=6$ superconformal Chern-Simons-matter theory with
$U(N)_k\times U(N)_{-k}$ gauge symmetry. In contrast to the
maximally supersymmetric $AdS_5\times S^5$ superbackground on
which propagates the IIB superstring involved into the
$AdS_5/CFT_4$ correspondence \cite{Maldacena} the
$AdS_4\times\mathbb{CP}^3$ superbackground preserves only 24 of 32
space-time supersymmetries that complicates the structure of the
superstring action \cite{GSWnew} and verification of the duality
conjecture (for recent review see, e.g. \cite{Klose}).

However, within the $AdS_4\times\mathbb{CP}^3$ superspace there exists a $(10|24)-$dimensional subsuperspace isomorphic to the $OSp(4|6)/(SO(1,3)\times U(3))$ supercoset manifold on which the $2d$ sigma-model can be constructed \cite{AF08}, \cite{Stefanski} using the supercoset approach elaborated to obtain the $AdS_5\times S^5$ superstring action \cite{MT98}-\cite{RoibanSiegel} \footnote{Another approach to the description of $OSp(4|6)/(SO(1,3)\times U(3))$ sigma-model based on introduction of the pure spinor variables was considered in \cite{PS}.}. One of important properties of such a $OSp(4|6)/(SO(1,3)\times U(3))$ sigma-model is that its equations of motion are manifestly classically integrable \cite{AF08}, \cite{Stefanski} similarly to the $AdS_5\times S^5$ superstring case \cite{BPR}. The $OSp(4|6)/(SO(1,3)\times U(3))$ sigma-model action is known to arise from the complete $AdS_4\times\mathbb{CP}^3$ superstring action \cite{GSWnew} when 8 fermionic coordinates associated with the broken supersymmetries are gauged away. As usual when imposing the gauge conditions on the level of the action non-trivial equations of motions for the gauged out variables should not be lost. It was proved in \cite{SW} that to leading order in the fermionic coordinates parametrizing the $OSp(4|6)/(SO(1,3)\times U(3))$ supermanifold corresponding equations of the $AdS_4\times\mathbb{CP}^3$ superstring constitute a subset of fermionic equations of the $OSp(4|6)/(SO(1,3)\times U(3))$ sigma-model. Here using that the $OSp(4|6)/(SO(1,3)\times U(3))$ sigma-model equations are formulated in terms of the $osp(4|6)$ Cartan forms we give a parametrization independent proof i.e. to all orders in the fermionic coordinates.

Organization of this note is the following. After recalling the $\mathbb{Z}_4-$grading of the $osp(4|6)$ superalgebra that is relevant to constructing geometric constituents of the $OSp(4|6)/(SO(1,3)\times U(3))$ superspace we prove that equations of motion corresponding to variation of the action of massless $AdS_4\times\mathbb{CP}^3$ superparticle on 8 fermionic superspace coordinates associated with the supersymmetries broken by the superbackground in the limit when those coordinates are put to zero are contained in the set of fermionic equations for the massless superparticle on the $OSp(4|6)/(SO(1,3)\times U(3))$ supermanifold. The proof is then generalized to the case of $AdS_4\times\mathbb{CP}^3$ superstring.

%%%%%%%%%%%%%%%%%%%%%%%%%%%%%%%%%%%%%%
%
\section{$osp(4|6)$ superalgebra and $OSp(4|6)/(SO(1,3)\times U(3))$ supermanifold}\label{sec2}
%
%%%%%%%%%%%%%%%%%%%%%%%%%%%%%%%%%%%%%%

At the heart of group-theoretic approach to description of geometry of symmetric coset spaces $G/H$ lies identification of the $g/h$ coset Cartan forms with the vielbein components and of the stability algebra $h$ Cartan forms with the connection 1-form of the manifold $G/H$. For the supercoset manifolds like $OSp(4|6)/(SO(1,3)\times U(3))$ one on which the classically integrable sigma-models can be defined \cite{BPR}, \cite{Polyakov}-\cite{Zarembo} important role is played by $\mathbb{Z}_4$ automorphism of the isometry superalgebra $\mathfrak g$ that generalizes $\mathbb{Z}_2$ automorphisms of the isometry algebras of corresponding bosonic coset manifolds. So that the (anti)commutation relations of $\mathfrak g$ can be cast into the $\mathbb{Z}_4-$graded form
\beq
[\mathfrak g_{(j)},\mathfrak g_{(k)}\}=\mathfrak g_{(j+k)mod4}
\eeq
with all the generators divided into 4 eigenspaces according to their $\mathbb{Z}_4$ eigenvalues
\beq
\Upsilon(\mathfrak g_{(k)})=i^k\mathfrak g_{(k)},\ k=0,1,2,3.
\eeq

For the $osp(4|6)$ superalgebra relevant $\mathbb{Z}_2-$graded
form of $so(2,3)\sim sp(4)$ and $su(4)\sim so(6)$ isometry
algebras of $AdS_4=SO(2,3)/SO(1,3)$ and $\mathbb{CP}^3=SU(4)/U(3)$
manifolds is given by the relations
\beq\label{ads4alg}
\begin{array}{rl}
\left[M_{0'm'},M_{0'n'}\right]=&M_{m'n'},\\
\left[M_{m'n'},M_{0'k'}\right]=&\eta_{n'k'}M_{0'm'}-\eta_{m'k'}M_{0'n'},\\
\left[M_{k'l'},M_{m'n'}\right]=&\eta_{k'n'}M_{l'm'}-\eta_{k'm'}M_{l'n'}\\
-&\eta_{l'n'}M_{k'm'}+\eta_{l'm'}M_{k'n'}.
\end{array}
\eeq
and
\begin{equation}\label{su4alg}
\begin{array}{rl}
[T_a,T^b]=&i(V_a{}^b+\delta_a^b V_c{}^c),\\[0.1cm]
[T_a,V_b{}^c]=&-i\delta_a^cT_b,\
[T^a,V_b{}^c]=i\delta_b^aT^c,\\[0.1cm]
[V_a{}^b,V_c{}^d]=&i(\delta^b_cV_a{}^d-\delta^d_aV_c{}^b).
\end{array}
\end{equation}
The $so(2,3)$ generators $M_{0'm'}$ and $M_{m'n'}$ ($m',n'=0,...,3$) are related to the $D=3$ conformal group generators as
\beq
\begin{array}{l}
M_{0'm}=\frac12(P_m+K_m),\ M_{0'3}=-D,\\[0.2cm] M_{3m}=\frac12(K_m-P_m).
\end{array}
\eeq
Thus the $\mathfrak g_{(0)}$
eigenspace is spanned by the generators $M_{m'n'}$ and $V_a{}^b$,
while $\mathfrak g_{(2)}$ by the generators $M_{0'm'}$ and $T_a$, $T^a$.
Using the isomorphism between the $osp(4|6)$ superalgebra and
$D=3$ $\mathcal N=6$ superconformal algebra that is the symmetry algebra of the ABJM gauge theory \cite{BLS} one can span $\mathfrak g_{(1)}$ and
$\mathfrak g_{(3)}$ eigenspaces by
\beq\label{gradedfermidef}
\begin{array}{c}
Q^{\vp{a}}_{(1)}\vp{Q}^a_\mu=Q^a_\mu+iS^a_\mu,\quad\bar Q_{(1)}\vp{Q}_{\mu a}=\bar Q_{\mu a}-i\bar S_{\mu a};\\
Q^{\vp{a}}_{(3)}\vp{Q}^a_\mu=Q^a_\mu-iS^a_\mu,\quad\bar Q_{(3)}\vp{Q}_{\mu a}=\bar Q_{\mu a}+i\bar S_{\mu a},
\end{array}
\eeq
where $Q^a_\mu$, $\bar Q_{\mu a}$ and $S^a_\mu$, $\bar S_{\mu a}$ are the generators of $D=3$ $\mathcal N=6$ super-Poincare and superconformal symmetries carrying $SL(2,\mathbb{R})$ spinor index $\mu=1,2$ and $SU(3)$ (anti)fundamental representation index $a=1,2,3$ in accordance with the decomposition of $SO(6)$ vector on the $SU(3)$ representations $\mathbf 6=\mathbf 3\oplus\bar{\mathbf 3}$.

Left-invariant $osp(4|6)$ Cartan forms in conformal basis are defined by the relation \cite{U08}
\beq
\begin{array}{rl}
\mathcal C(d)=&\mathscr G^{-1}d\mathscr G=\omega^m(d)P_m+c^m(d)K_m\\[0.2cm]
+&\Delta(d)D+\Omega_a(d)T^a+\Omega^a(d)T_a\\[0.2cm]
+&\omega^\mu_a(d)Q^a_\mu+\bar\omega^{\mu a}(d)\bar Q_{\mu a}+\chi_{\mu
a}(d)S^{\mu a}\\[0.2cm]
+&\bar\chi^a_\mu(d)\bar S^\mu_a+G^{mn}(d)M_{mn}+\Omega_a{}^b(d)V_b{}^a.
\end{array}
\eeq
By analogy with the $osp(4|6)$ generators they can be arranged in the form
\beq\label{gradedcf}
\mathcal C(d)=\mathcal C_{(0)}+\mathcal C_{(2)}+\mathcal C_{(1)}+\mathcal C_{(3)},
\eeq
where each summand takes value in respective eigenspace under the $\mathbb{Z}_4$ grading
\beq
\begin{array}{rl}
\mathcal C_{(0)}=&2G^{3m}M_{3m}+G^{mn}M_{mn}+\Omega_a{}^bV_b{}^a,\\[0.2cm]
\mathcal C_{(2)}=&2G^{0'm}M_{0'm}+\Delta D+\Omega_aT^a+\Omega^aT_a,\\[0.2cm]
\mathcal C_{(1)}=&\omega_{(1)}\vp{\omega}^\mu_{a}Q_{(1)}\vp{Q}^a_\mu+\bar\omega^{\vp{\mu c}}_{(1)}\vp{\bar\omega}^{\mu a}\bar Q_{(1)\mu a},\\[0.2cm]
\mathcal C_{(3)}=&\omega_{(3)}\vp{\omega}^\mu_{a}Q_{(3)}\vp{Q}^a_\mu+\bar\omega^{\vp{\mu c}}_{(3)}\vp{\bar\omega}^{\mu a}\bar Q_{(3)\mu a}.
\end{array}
\eeq
So that the Cartan forms $G^{0'm}(d)=\frac12(\omega^m(d)+c^m(d))$, $\Delta(d)$ and $\Omega_a(d)$, $\Omega^a(d)$ associated with the $\mathfrak g_{(2)}$ generators are identified with the $OSp(4|6)/(SO(1,3)\times U(3))$ supervielbein bosonic components and
fermionic Cartan forms
\beq
\begin{array}{c}
\omega_{(1)}\vp{\omega}^\mu_{a}(d)=\frac12(\omega^\mu_a(d)+i\chi^\mu_a(d)),\\[0.2cm]
\omega_{(3)}\vp{\omega}^\mu_{a}(d)=\frac12(\omega^\mu_a(d)-i\chi^\mu_a(d))
\end{array}
\eeq
and c.c. are identified with the $OSp(4|6)/(SO(1,3)\times U(3))$ supervielbein fermionic components. Accordingly Cartan forms for the $\mathfrak g_{(0)}$ generators
$G^{3m}(d)=-\frac12(\omega^m(d)-c^m(d))$, $G^{mn}(d)$ and $\Omega_a{}^b(d)$
are identified with the connection 1-form on the $OSp(4|6)/(SO(1,3)\times U(3))$ supermanifold.

\section{$OSp(4|6)/(SO(1,3)\times U(3))$ superparticle}

Massless superparticle action on the $OSp(4|6)/$\\ $(SO(1,3)\times U(3))$ supercoset manifold \cite{Stefanski}
\beq\label{cosetaction}
\mathscr S_{\mbox{\scriptsize{sp}}}=\int\frac{d\tau}{e}(G_{\tau}\vp{G}^{0'}\vp{G}_{m}G_\tau\vp{G}^{0'm}+\Delta_\tau\Delta_\tau+\Omega_{\tau a}\Omega_\tau{}^a),
\eeq
constructed out of the world-line pullbacks of Cartan forms from the $\mathcal C_{(2)}$ eigenspace, is by definition invariant under the $OSp(4|6)$ global symmetry acting as the left group multiplication on $\mathscr G\in OSp(4|6)/(SO(1,3)\times U(3))$. Action variation w.r.t. Lagrange multiplier $e(\tau)$ produces the mass-shell constraint
\beq\label{spms}
G_{\tau}\vp{G}^{0'}\vp{G}_{m}G_\tau\vp{G}^{0'm}+\Delta_\tau\Delta_\tau+\Omega_{\tau a}\Omega_\tau{}^a=0.
\eeq
Applicability of the supercoset approach implies that $\Omega_{\tau a}\not=0$ \cite{AF08}, \cite{GSWnew} so that $
G_{\tau}\vp{G}^{0'}\vp{G}_{m}G_\tau\vp{G}^{0'm}+\Delta_\tau\Delta_\tau\equiv G_\tau\cdot G_\tau+\Delta^2_\tau\not=0$.

Variation of the action (\ref{cosetaction}) is contributed by variations of the constituent $so(2,3)/so(1,3)\oplus su(4)/u(3)$ Cartan forms that can be obtained from the general expression
\beq
\delta\mathcal F(d)=di_\delta\mathcal F+i_\delta d\mathcal F
\eeq
upon substitution in the second summand Maurer-Cartan equations for $\mathcal C_{(2)}$ Cartan forms \cite{U08}
\beq
\begin{array}{rl}
dG^{0'm}-&2G^{3m}\wedge\Delta-2G^{mn}\wedge G^{0'}\vp{G}_n\\[0.2cm]
+&2i\omega_{(1)}\vp{\omega}^\mu_{a}\wedge\sigma^m_{\mu\nu}\bar\omega_{(1)}\vp{\bar\omega}^{\nu a}
+(1\leftrightarrow3)=0,\\[0.2cm]
d\Delta+&2G^{3m}\wedge G^{0'}\vp{G}_m\\[0.2cm]
+&2\omega_{(1)}\vp{\omega}^\mu_{a}\wedge\bar\omega_{(1)}\vp{\bar\omega}_\mu^{a}-(1\leftrightarrow3)=0,\\[0.2cm]
d\Omega^a+&i\Omega^b\wedge(\Omega_b{}^a+\delta^a_b\Omega_c{}^c)\\[0.2cm]
-&2i\varepsilon^{abc}\omega_{(1)}\vp{\omega}^\mu_{b}\wedge\omega_{(1)\mu c}
+(1\leftrightarrow3)=0,\\[0.2cm]
d\Omega_a+&i(\Omega_a{}^b+\delta^b_a\Omega_c{}^c)\wedge\Omega_b\\[0.2cm]
-&2i\varepsilon_{abc}\bar\omega_{(1)}\vp{\omega}^{\mu b}\!\wedge\bar\omega_{(1)}\vp{\omega}_{\mu}\vp{\omega}^{c}\!
+(1\leftrightarrow3)=0.
\end{array}
\eeq
Then taking the contractions of the Cartan forms (\ref{gradedcf}) with the variation symbol $i_\delta$ as independent parameters yields the set of superparticle equations of motion \cite{U12-2}.

Here we concentrate on the fermionic equations that can be brought to the form
\begin{equation}\label{speom1}
\mathscr M_{(1)}\vp{\mathscr M}^{\vp{\mu}}_{\vp{\hat a}\tau}\vp{\mathscr M}_{\hat a}^\mu\vp{\mathscr M}_{\vp{\hat a}\nu}^{\vp{\mu}\hat b}\left(
\begin{array}{c}
\omega_{(1)}\vp{\omega}^{\vp{\nu}}_{\vp{b}\tau}\vp{\omega}^\nu_b \\[0.1cm]
\bar\omega_{(1)}\vp{\omega}_\tau\vp{\omega}^{\nu b}
\end{array}\right)=0
\end{equation}
where the $12\times12$ matrix $\mathscr M_{(1)}\vp{\mathscr M}^{\vp{\mu}}_{\vp{\hat a}\tau}\vp{\mathscr M}_{\hat a}^\mu\vp{\mathscr M}_{\vp{\hat a}\nu}^{\vp{\mu}\hat b}$ of rank 8 equals
\begin{equation}
\left(
\begin{array}{cc}
\delta_a^bm_{\tau}\vp{m}^\mu\vp{m}_\nu & -i\delta^\mu_\nu\varepsilon_{acb}\Omega_\tau{}^c \\[0.2cm]
-i\delta^\mu_\nu\varepsilon^{acb}\Omega_{\tau c} & -\delta^a_b\bar m_{\tau}\vp{m}^\mu\vp{m}_\nu
\end{array}
\right)
\end{equation}
and $2\times 2$ matrices
\begin{equation}
\begin{array}{rcl}
m_{\tau}\vp{m}^\mu\vp{m}_\nu&=&-iG_\tau{}^{0'm}\sigma_m{}^\mu{}_\nu+\Delta_\tau\delta^\mu_\nu,\\
\bar
m_\tau\vp{m}^\mu\vp{m}_\nu&=&iG_\tau{}^{0'm}\sigma_m{}^\mu{}_\nu+\Delta_\tau\delta^\mu_\nu
\end{array}
\end{equation}
obey the relation
\begin{equation}\label{rel1}
m_{\tau}\vp{m}^\mu\vp{m}_\nu\bar m_{\tau}\vp{m}^\nu\vp{m}_\lambda=\delta^\mu_\lambda(G_\tau\cdot G_\tau+\Delta^2_\tau).
\end{equation}
Similarly equations for Cartan forms corresponding to the generators from the $\mathfrak g_{(3)}$ eigenspace read
\begin{equation}\label{speom3}
\mathscr M_{(3)}\vp{\mathscr M}^{\vp{\mu}}_{\vp{\hat a}\tau}\vp{\mathscr M}_{\hat a}^\mu\vp{\mathscr M}_{\vp{\hat a}\nu}^{\vp{\mu}\hat b}\left(
\begin{array}{c}
\omega_{(3)}\vp{\omega}^{\vp{\nu}}_{\vp{b}\tau}\vp{\omega}^\nu_b \\[0.1cm]
\bar\omega_{(3)}\vp{\omega}_\tau\vp{\omega}^{\nu b}
\end{array}\right)=0
\end{equation}
with
\begin{equation}
\mathscr M_{(3)}\vp{\mathscr M}^{\vp{\mu}}_{\vp{\hat a}\tau}\vp{\mathscr M}_{\hat a}^\mu\vp{\mathscr M}_{\vp{\hat a}\nu}^{\vp{\mu}\hat b}=
\left(
\begin{array}{cc}
-\delta_a^b\bar m_{\tau}\vp{m}^\mu\vp{m}_\nu & i\delta^\mu_\nu\varepsilon_{acb}\Omega_\tau{}^c \\[0.2cm]
i\delta^\mu_\nu\varepsilon^{acb}\Omega_{\tau c} & \delta^a_bm_{\tau}\vp{m}^\mu\vp{m}_\nu
\end{array}
\right).
\end{equation}

At the same time since the $OSp(4|6)/(SO(1,3)\times U(3))$ superparticle action comes about upon gauging away 8 coordinates from the broken supersymmetries sector in the action for massless superparticle on the $AdS_4\times\mathbb{CP}^3$ superbackground \cite{U12-2} that can be viewed as the zero-mode
limit of the $AdS_4\times\mathbb{CP}^3$ superstring \cite{GSWnew} or the mass-to-zero limit of the D0-brane \cite {GrSW} the equations of motion for gauged away fermions require special attention. It appears that they are non-trivial and can be brought to the form \cite{U12-2}
\beq\label{spgaugedeom}
\Omega_\tau{}^b\omega_{(1,3)}\vp{\omega}^{\vp{\nu}}_{\vp{b}\tau}\vp{\omega}^\nu_b=\Omega_{\tau
b}\bar\omega_{(1,3)}\vp{\omega}_\tau\vp{\omega}^{\nu b}=0.
\eeq
Below we shall show that 8 equations (\ref{spgaugedeom}) are contained in the set of equations (\ref{speom1}), (\ref{speom3}).

Consider in detail the system of equations (\ref{speom1}). Applying rank 8 projection matrix
\begin{equation}
\Pi_{(1)}\!=\!\!\left(\!\!\!
\begin{array}{cc}
\delta_a^b\delta^\mu_\nu & \hspace*{-0.4cm}-i\frac{m_{\tau}\vp{m}^\mu\vp{m}_\nu}{G_\tau\cdot G_\tau+\Delta^2_\tau}\varepsilon_{acb}\Omega_\tau{}^c \\
i\frac{\bar m_{\tau}\vp{m}^\mu\vp{m}_\nu}{G_\tau\cdot
G_\tau+\Delta^2_\tau}\varepsilon^{acb}\Omega_{\tau c} &
\hspace*{-0.4cm}\delta^a_b\delta^\mu_\nu
\end{array}
\!\!\!\right)
\end{equation}
that satisfies $\Pi^3_{(1)}-3\Pi^2_{(1)}+2\Pi_{(1)}=0$ and recalling that $G_\tau\cdot G_\tau+\Delta^2_\tau\not=0$ allows to bring (\ref{speom1}) to the block-diagonal form
\begin{equation}\label{speom1'}
%\frac{1}{G_\tau\cdot G_\tau+\Delta^2_\tau}
\left(\!\!
\begin{array}{cc}
m_{\tau}\vp{m}^\mu\vp{m}_\nu\Omega_{\tau a}\Omega_\tau{}^b & \hspace*{-0.3cm}0 \\[0.2cm]
0 & \hspace*{-0.3cm}-\bar m_{\tau}\vp{m}^\mu\vp{m}_\nu\Omega_\tau{}^a\Omega_{\tau b}
\end{array}
\!\!\right)\!\left(\!\!
\begin{array}{c}
\omega_{(1)}\vp{\omega}^{\vp{\nu}}_{\vp{b}\tau}\vp{\omega}^\nu_b \\[0.2cm]
\bar\omega_{(1)}\vp{\omega}_\tau\vp{\omega}^{\nu b}
\end{array}\!\!\right)\!\!=\!0.
\end{equation}
Further using (\ref{rel1}) yields that the system (\ref{speom1'}) is equivalent to 4 equations
\begin{equation}
\Omega_\tau{}^b\omega_{(1)}\vp{\omega}^{\vp{\nu}}_{\vp{b}\tau}\vp{\omega}^\nu_b=\Omega_{\tau
b}\bar\omega_{(1)}\vp{\omega}_\tau\vp{\omega}^{\nu b}=0
\end{equation}
coinciding with (\ref{spgaugedeom}). Analogously it can be shown that the system (\ref{speom3}) includes equations
\begin{equation}
\Omega_\tau{}^b\omega_{(3)}\vp{\omega}^{\vp{\nu}}_{\vp{b}\tau}\vp{\omega}^\nu_b=\Omega_{\tau
b}\bar\omega_{(3)}\vp{\omega}_\tau\vp{\omega}^{\nu b}=0.
\end{equation}

\section{$OSp(4|6)/(SO(1,3)\times U(3))$ sigma-model}

The $OSp(4|6)/(SO(1,3)\times U(3))$ sigma-model \cite{AF08}, \cite{Stefanski} action in conformal basis for the $osp(4|6)/(so(1,3)\times u(3))$ Cartan forms can be written as \cite{U08}
\beq
\mathscr S_{\mathrm{s-m}}=\int d^2\xi(\mathscr L_{\mathrm{kin}}+\mathscr L_{\mathrm{WZ}})
\eeq
with the kinetic and Wess-Zumino Lagrangians given by
\beq
\begin{array}{rl}
\mathscr L_{\mathrm{kin}}=&-\frac12\gamma^{ij}\!\left(\! G^{0'm}_iG_{j}{}^{0'}{}_{m}\!+\!\Delta_i\Delta_j\!+\!\Omega_{ia}\Omega_{j}{}^a\!\right)\\[0.2cm]
=&-\frac12\gamma^{ij}\left(\frac14(\omega_{im}+ c_{im})(\omega^m_j+ c^m_j)\right.\\[0.2cm]
+&\left.\Delta_i\Delta_j+\Omega_{ia}\Omega_{j}{}^a\right)
\end{array}
\eeq
and
\beq
\begin{array}{rl}
\mathscr L_{\mathrm{WZ}}=&-\varepsilon^{ij}\!\left(\omega_{(1)}\vp{\omega}^{\vp{\mu}}_i\vp{\omega}^\mu_{\vp{i}a}\varepsilon_{\mu\nu}\bar\omega_{(3)j}\vp{\bar\omega}^{\nu a}+(1\leftrightarrow3)\right)\\[0.2cm]
=&-\frac{1}{2}\varepsilon^{ij}\left(\omega^{\vp{\mu}}_i\vp{\omega}^\mu_{\vp{i}a}\varepsilon_{\mu\nu}\bar\omega^{\vp{\mu}}_{j}\vp{\bar\omega}^{\nu
a}+\chi_{i\mu
a}\varepsilon^{\mu\nu}\bar\chi^{\vp{\nu}}_j\vp{\bar\chi}^a_{\nu}\right).
\end{array}
\eeq

Fermionic equations of the $OSp(4|6)/(SO(1,3)\times U(3))$ sigma-model \cite{AF08}, \cite{Stefanski} can be cast into the following form \cite{U08} generalizing superparticle equations (\ref{speom1}), (\ref{speom3})
\begin{equation}\label{smeom1}
V^{ij}_+\mathscr M_{(1)}\vp{\mathscr M}^{\vp{\mu}}_{\vp{\hat a}i}\vp{\mathscr M}_{\hat a}^\mu\vp{\mathscr M}_{\vp{\hat a}\nu}^{\vp{\mu}\hat b}\left(
\begin{array}{c}
\omega_{(1)}\vp{\omega}^{\vp{\nu}}_{\vp{b}j}\vp{\omega}^\nu_b \\[0.1cm]
\bar\omega_{(1)}\vp{\omega}_j\vp{\omega}^{\nu b}
\end{array}\right)=0
\end{equation}
and
\begin{equation}\label{smeom3}
V^{ij}_-\mathscr M_{(3)}\vp{\mathscr M}^{\vp{\mu}}_{\vp{\hat a}i}\vp{\mathscr M}_{\hat a}^\mu\vp{\mathscr M}_{\vp{\hat a}\nu}^{\vp{\mu}\hat b}\left(
\begin{array}{c}
\omega_{(3)}\vp{\omega}^{\vp{\nu}}_{\vp{b}j}\vp{\omega}^\nu_b \\[0.1cm]
\bar\omega_{(3)}\vp{\omega}_j\vp{\omega}^{\nu b}
\end{array}\right)=0,
\end{equation}
where $V^{ij}_{\pm}=\frac12(\gamma^{ij}\pm\varepsilon^{ij})$ are (anti)self-dual world-sheet projectors and $12\times 12$ matrices that now also bear world-sheet vector index $i=(\tau, \sigma)$ are defined as
\begin{equation}
\mathscr M_{(1)}\vp{\mathscr M}^{\vp{\mu}}_{\vp{\hat a}i}\vp{\mathscr M}_{\hat a}^\mu\vp{\mathscr M}_{\vp{\hat a}\nu}^{\vp{\mu}\hat b}=
\left(
\begin{array}{cc}
\delta_a^bm_{i}\vp{m}^\mu\vp{m}_\nu & -i\delta^\mu_\nu\varepsilon_{acb}\Omega_i{}^c \\[0.2cm]
-i\delta^\mu_\nu\varepsilon^{acb}\Omega_{ic} & -\delta^a_b\bar m_{i}\vp{m}^\mu\vp{m}_\nu
\end{array}
\right)
\end{equation}
and
\begin{equation}
\mathscr M_{(3)}\vp{\mathscr M}^{\vp{\mu}}_{\vp{\hat a}i}\vp{\mathscr M}_{\hat a}^\mu\vp{\mathscr M}_{\vp{\hat a}\nu}^{\vp{\mu}\hat b}=
\left(
\begin{array}{cc}
-\delta_a^b\bar m_{i}\vp{m}^\mu\vp{m}_\nu & i\delta^\mu_\nu\varepsilon_{acb}\Omega_i{}^c \\[0.2cm]
i\delta^\mu_\nu\varepsilon^{acb}\Omega_{ic} & \delta^a_bm_{i}\vp{m}^\mu\vp{m}_\nu
\end{array}
\right).
\end{equation}

Apart from these equations similarly to the superparticle case there remain non-trivial equations for 8 fermions from the sector of broken supersymmetries that can be brought to the form \cite{U12-1}
\begin{equation}\label{smgaugedeom}
\begin{array}{c}
V^{ij}_+\Omega_i{}^b\omega_{(1)}\vp{\omega}^{\vp{\nu}}_{\vp{b}j}\vp{\omega}^\nu_b=V^{ij}_+\Omega_{ib}\bar\omega_{(1)}\vp{\omega}_j\vp{\omega}^{\nu b}=0,\\[0.2cm]
V^{ij}_-\Omega_i{}^b\omega_{(3)}\vp{\omega}^{\vp{\nu}}_{\vp{b}j}\vp{\omega}^\nu_b=V^{ij}_-\Omega_{ib}\bar\omega_{(3)}\vp{\omega}_j\vp{\omega}^{\nu b}=0.
\end{array}
\end{equation}
In the remaining part of this section we shall show that they are contained in the set of fermionic equations (\ref{smeom1}), (\ref{smeom3}) of the $OSp(4|6)/(SO(1,3)\times U(3))$ sigma-model.

Recalling that the action of the projectors $V^{ij}_{\pm}$ on a vector can be factorized as
\begin{equation}
\begin{array}{c}
V^{ij}_{\pm}F_j=V^i_{\pm}F_{\mp}:\quad
V^i_{\pm}=\frac12\left(\begin{array}{c}
1\\
\frac{\gamma^{\tau\sigma}\mp1}{\gamma^{\tau\tau}}
\end{array}\right),\\[0.3cm]
F_{\mp}=\gamma^{\tau\tau}F_\tau+(\gamma^{\tau\sigma}\pm1)F_\sigma
\end{array}
\end{equation}
one can write Eqs.~(\ref{smeom1}), (\ref{smeom3}) in the form
\begin{equation}\label{smeom1'}
\mathscr M_{(1)}\vp{\mathscr M}^{\vp{\mu}}_{\vp{\hat a}+}\vp{\mathscr M}_{\hat a}^\mu\vp{\mathscr M}_{\vp{\hat a}\nu}^{\vp{\mu}\hat b}\left(
\begin{array}{c}
\omega_{(1)}\vp{\omega}^{\vp{\nu}}_{\vp{b}-}\vp{\omega}^\nu_b \\[0.1cm]
\bar\omega_{(1)}\vp{\omega}_-\vp{\omega}^{\nu b}
\end{array}\right)=0
\end{equation}
and
\begin{equation}\label{smeom3'}
\mathscr M_{(3)}\vp{\mathscr M}^{\vp{\mu}}_{\vp{\hat a}-}\vp{\mathscr M}_{\hat a}^\mu\vp{\mathscr M}_{\vp{\hat a}\nu}^{\vp{\mu}\hat b}\left(
\begin{array}{c}
\omega_{(3)}\vp{\omega}^{\vp{\nu}}_{\vp{b}+}\vp{\omega}^\nu_b \\[0.1cm]
\bar\omega_{(3)}\vp{\omega}_+\vp{\omega}^{\nu b}
\end{array}\right)=0,
\end{equation}
where now
\begin{equation}
\mathscr M_{(1)}\vp{\mathscr M}^{\vp{\mu}}_{\vp{\hat a}+}\vp{\mathscr M}_{\hat a}^\mu\vp{\mathscr M}_{\vp{\hat a}\nu}^{\vp{\mu}\hat b}=
\left(
\begin{array}{cc}
\delta_a^bm_{+}\vp{m}^\mu\vp{m}_\nu & -i\delta^\mu_\nu\varepsilon_{acb}\Omega_{+}{}^c \\[0.2cm]
-i\delta^\mu_\nu\varepsilon^{acb}\Omega_{+c} & -\delta^a_b\bar m_{+}\vp{m}^\mu\vp{m}_\nu
\end{array}
\right)
\end{equation}
and
\begin{equation}
\mathscr M_{(3)}\vp{\mathscr M}^{\vp{\mu}}_{\vp{\hat a}-}\vp{\mathscr M}_{\hat a}^\mu\vp{\mathscr M}_{\vp{\hat a}\nu}^{\vp{\mu}\hat b}=
\left(
\begin{array}{cc}
-\delta_a^b\bar m_{-}\vp{m}^\mu\vp{m}_\nu & i\delta^\mu_\nu\varepsilon_{acb}\Omega_{-}{}^c \\[0.2cm]
i\delta^\mu_\nu\varepsilon^{acb}\Omega_{-c} & \delta^a_bm_{-}\vp{m}^\mu\vp{m}_\nu
\end{array}
\right),
\end{equation}
that is similar to the superparticle equations (\ref{speom1}), (\ref{speom3}). Note that application of the projectors $V^{ij}_{\pm}$ to the Virasoro constraints gives \cite{U08}
\begin{equation}
G_{\pm}{}^{0'}{}_mG_{\pm}{}^{0'm}+\Delta^2_{\pm}+\Omega_{\pm
a}\Omega_{\pm}{}^a=0
\end{equation}
generalizing the superparticle mass-shell condition (\ref{spms}) so that the matrices $m_{\pm}\vp{m}^\mu\vp{m}_\nu$ can be shown to satisfy the relations
\begin{equation}
m_{\pm}\vp{m}^\mu\vp{m}_\nu\bar m_{\pm}\vp{m}^\nu\vp{m}_\lambda=\delta^\mu_\lambda(G_{\pm}\cdot G_{\pm}+\Delta^2_{\pm})
\end{equation}
generalizing (\ref{rel1}). Thus repeating the analysis performed for the superparticle case we conclude that Eqs.~(\ref{smgaugedeom}) are contained in the system of $OSp(4|6)/(SO(1,3)\times U(3))$ sigma-model fermionic equations (\ref{smeom1}), (\ref{smeom3}).

%%%%%%%%%%%%%%%%%%%%%%%%%%%%%%%%
\section{Conclusion}
%%%%%%%%%%%%%%%%%%%%%%%%%%%%%%%%
We have proved that in the partial $\kappa-$symmetry gauge in which 8 fermionic coordinates corresponding to the supersymmetries broken by $AdS_4\times\mathbb{CP}^3$ superbackground are set to zero equations of motion obtained by variation of the $AdS_4\times\mathbb{CP}^3$ superstring action w.r.t to such coordinates are contained in the fermionic equations of the $OSp(4|6)/(SO(1,3)\times U(3))$ sigma-model generalizing the argument of Ref.~\cite{SW}. This provides yet another necessary consistency check of the $OSp(4|6)/(SO(1,3)\times U(3))$ sigma-model approach to the description of the certain sector of $AdS_4\times\mathbb{CP}^3$ superstring dynamics and shows that in such a partial $\kappa-$symmetry gauge all the non-trivial equations of $AdS_4\times\mathbb{CP}^3$ superstring can be reduced to those derivable from the $OSp(4|6)/(SO(1,3)\times U(3))$ sigma-model action.

The situation changes when the gauge condition is relaxed to allow non-trivial dynamics in the sector of broken supersymmetries. This lifts restrictions on admissible string motions imposed in the framework of the $OSp(4|6)/(SO(1,3)\times U(3))$ sigma-model approach and also equations of motion for the fermions related to the broken supersymmetries become independent of those for the fermions associated with the unbroken supersymmetries of the background. Both sets are involved in proving integrability of the $AdS_4\times\mathbb{CP}^3$ superstring/superparticle equations beyond the $OSp(4|6)/(SO(1,3)\times U(3))$ supercoset \cite{SW}, \cite{CSW}\footnote{Generalizations to the superstring models on other superbackgrounds part of which admits supercoset description were examined in \cite{STWZ}, \cite{CSW}, \cite{Sundin}.}, \cite{U12-1}, \cite{U12-2}.

%%%%%%%%%%%%%%%%%%%%%%%%%%%%%%%%
\section*{Acknowledgement}
The author is grateful to A.A.~Zheltukhin for interesting discussions.
%%%%%%%%%%%%%%%%%%%%%%%%%%%%%%%%

\end{multicols}
\end{document}